\documentclass[twocolumn,twoside,slac_two]{revtex4}
\usepackage{graphicx}
\usepackage{fancyhdr}
\begin{document}

\title{Prospects for Measuring the Cosmic-ray Proton Spectrum Using the LAT Instrument on the Fermi Gamma-ray Space Telescope}

%

\author{P. D. Smith, R. E. Hughes, B. L. Winer, T. W. Wood for the Fermi LAT Collaboration}
\affiliation{Center for Cosmology and Astro-Particle Physics, Department of Physics, The Ohio State University, Columbus, OH 43210, USA}

\begin{abstract}
The Fermi Gamma-Ray Space Telescope was launched in June 2008 and the onboard Large Area Telescope (LAT) has been collecting data since August of that same year. The LAT is currently being used to study a wide range of science topics in high-energy astrophysics, one of which is the study of high-energy cosmic rays. The LAT has recently demonstrated its ability to measure cosmic-ray electrons, and the Fermi LAT Collaboration has published a measurement of the high-energy cosmic-ray electron spectrum in the 20 GeV to 1 TeV energy range. This talk will discuss the prospects for using the LAT to perform a similar analysis to measure cosmic-ray proton events. The instrument response for cosmic-ray protons will be characterized and an assessment of the potential to measure the cosmic-ray proton energy spectrum will be presented. 
\end{abstract}

\maketitle

\thispagestyle{fancy}


\section{Introduction}
The Fermi Gamma-ray Space Telescope was designed for the study of many interesting topics in astrophysics ranging from pulsars, active galactic nuclei, gamma-ray bursts, and the indirect detection of dark matter.  Another intriguing subject that has attracted strong interest is the use of the Fermi LAT in the study of cosmic rays.  Early in the design stages of the LAT, the potential for making a cosmic-ray electron measurement was acknowledged \cite{Ormes}, \cite{Moiseev}. Recently the Fermi LAT Collaboration demonstrated this ability and published a high-statistics measurement of the cosmic-ray electron spectrum in the energy range from 20 GeV to 1 TeV \cite{CREpaper}, containing about 4.5 million events collected over a six month period from August 2008 to January 2009.  The electron spectrum measurement can be used in the study of cosmic-ray propagation and in the constraint of the diffuse gamma-ray emission \cite{CRPropPaper}.  Furthermore, it may also be possible to use the LAT to obtain a measurement of the cosmic-ray proton spectrum, and studies are being conducted within the LAT collaboration to explore this possibility.  We present some preliminary findings illustrating the prospects for such an analysis.

First a general introduction to the Fermi LAT instrument will be given, followed by a brief description of the analysis performed for the electron spectrum measurement.  Then we will present a discussion of the prospects for a cosmic-ray proton spectrum analysis.

\section{The Fermi Gamma-ray Space Telescope}
The Fermi Gamma-ray Space Telescope was launched on June 11, 2008 from Kennedy Space Center.  The spacecraft maintains a circular orbit of 565 km, and it completes one orbit approximately every 90 minutes.  Fermi's orbit has an inclination of 25.6$^\circ$ and a precession period of 55 days.  The mission has a minimum livetime of 5 years, with a goal of operating for 10 years.

The Fermi satellite is composed of two instruments.  The main instrument is the Large Area Telescope (LAT), which is a pair conversion telescope sensitive to the energy range from 20 MeV up to greater than 300 GeV.  The second instrument onboard the satellite is the Gamma-ray Burst Monitor (GBM) and is designed to detect GRBs in the 8 keV to 40 MeV energy range.  It consists of 12 NaI and 2 BGO detectors positioned around the spacecraft, and has a large field-of-view allowing it to constantly monitor the entire unocculted sky.  The analyses described here pertain to the LAT instrument, and a brief description of this detector follows.

\subsection{The LAT Instrument}
The LAT is surrounded on the top and the sides by a segmented anti-coincidence detector (ACD) comprised of 89 scintillating tiles. These scintillating tiles have a radiation length of 4\% and an efficiency of 99.97\% for minimum ionizing particles.  Below the ACD are 16 identical tower modules arranged in a 4$\times$4 array.  Each tower is composed of, from top to bottom, a tracker, a calorimeter, and an electronics module for data acquisition.  The tracker consists of 18 double-planed layers of Silicon strip arrays with 16 Tungsten foils interleaved.  The top 12 layers have a thin foil (3\% R.L) while the next 4 layers have a thicker foil (18\% R.L).  The bottom two layers have no foil.  The calorimeter consists of CsI(Tl) scintillating crystals arranged into a hodoscopic array of 8 layers, with 12 crystals in each layer.  It has a total radiation length of 8.6 $X_\circ$ on-axis, and a total thickness of 71.8 g/cm$^2$.  A more detailed description of the LAT detector can be found in \cite{LATpaper}.

\begin{figure}[h]
\centering
\includegraphics[width=80mm]{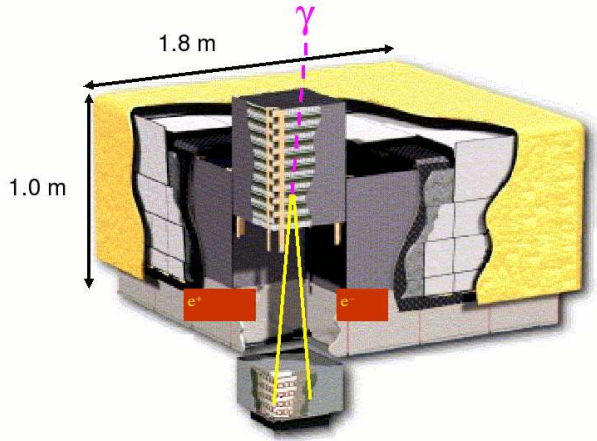}
\caption{The LAT instrument is a pair-conversion telescope designed to track the paths of an electron/positron pair resulting from a gamma-ray annihilation and to measure and image the resulting EM shower.} \label{LATschematic}
\end{figure}

The onboard event processing occurs first through a hardware trigger and is followed by a software filter.  The rate of events passing the trigger is between 2 to 4 kHz and has substantial variations due to orbital effects.  After the onboard software filter, the event rate is reduced to between 400 to 500 Hz, all of which is downlinked and undergoes further event processing on the ground.  For the gamma-ray analyses, the rate is further reduced to $\sim$ 1 Hz or less after the cosmic-ray background rejection analysis while still preserving a gamma-ray efficiency exceeding 75\%.  It is worth noting then for the purpose of the discussion here that the majority of the downlinked events are in fact cosmic-ray events.

\section{Overview of the HE Cosmic-ray Electron Spectrum Measurement}
Owing to its design as a pair conversion telescope for detecting gamma rays, the LAT is intrinsically capable of being used as an electron telescope.  In particular, the Silicon tracker was designed to track the paths of the $e^{+}$/$e^{-}$ pair resulting from a gamma-ray conversion, and the CsI(Tl) calorimeter was designed to measure their energy and to image the resulting EM shower.  Consequently, the photon event reconstruction algorithms also work well for an electron analysis.  Another advantage is that the ACD can be used to minimize the gamma-ray contamination in an electron event selection.  The main obstacle in event selection is instead in the rejection of hadrons.  In addition, the LAT is not able to distinguish between electrons and positrons, and thus the resulting measurement is actually a summed electron plus positron spectrum.  Henceforth, the term {\it electron} will simply be used to refer to the sum.

The events for this analysis were collected from the onboard software filter in a way as to minimize the bias from the filter, which is designed to reject charged particles.  This was achieved via a feature in the onboard software filter that automatically sends to ground any triggered event that deposits greater than 20 GeV in the calorimeter.  Next, in order to select the electron events, all three subsystems of the LAT were utilized (ACD, Tracker, and Calorimeter).  The initial event selection was based on various quantities derived from each system.  Next, two classification trees were developed, one using tracker related quantities and another that used variables from the calorimeter, each giving an event-based electron probability.  These probabilities were then combined in an energy dependent scheme to provide the final event selection that gave the required hadron rejection power.  The residual hadron contamination was then estimated from applying the event selections to Fermi LAT Monte Carlo simulations of cosmic-ray protons, which has a spectrum modelled from previous cosmic-ray experiments.  The rate of protons passing the electron selections, as a function of reconstructed energy, was then subtracted from the rate of electron candidate events.  The resulting reconstructed energy distribution was then used as input for an algorithm designed to calculate the estimated distribution of the incoming energy for these events, a process known as energy unfolding.  From this unfolded energy distribution, the cosmic-ray electron spectrum can be calculated by dividing by the livetime and the geometry factor.  A more detailed discussion of the analysis performed for the electron spectrum measurement can be found in \cite{CREpaper}.

\begin{figure}[h]
\centering 
\includegraphics[width=80mm]{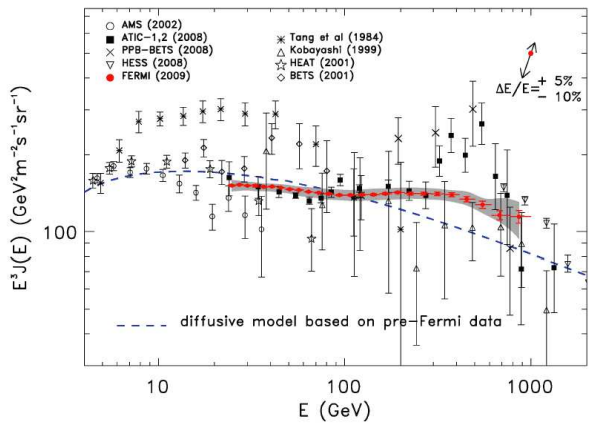}
\caption{The Fermi LAT cosmic-ray electron spectrum (shown in red) compared with other measurements.  The systematic errors are shown with the gray band.  The two-headed arrow in the top-right corner of the figure gives size and direction of the rigid shift of the spectrum implied by a shift of $^{+5\%}_{-10\%}$ of the absolute energy, corresponding to the present estimate of the uncertainty of the LAT energy scale.  A model based on pre-Fermi data is shown in the blue dashed line.} \label{CREspectrum}
\end{figure}

Figure \ref{CREspectrum} shows the published spectrum from the paper.  The errors in the spectrum are dominated by systematic uncertainties.  The spectrum is incompatible with a diffusive model based on pre-Fermi cosmic-ray data, shown by the dashed line in the figure, and in addition, no evidence of a prominent spectral feature is observed.

Future efforts will be to extend the energy range of the LAT's electron spectrum measurement and to increase the selection efficiency at energies close to 1 TeV.  Future analyses will also be aimed at searching for anisotropies in the arrival directions of the cosmic-ray electrons.

\section{Prospects for Measuring the Cosmic-ray Proton Spectrum}
We are currently exploring methods for producing a cosmic-ray proton spectrum measurement using the LAT instrument.  This analysis is inherently more challenging than the analysis for the electron spectrum measurement.  The main cause for this is that the thickness of the calorimeter is optimized for electromagnetic showers rather than hadronic showers.  Whereas the calorimeter is 8.6 $X_\circ$ on-axis for EM showers, it is only about 0.43 interaction lengths for hadronic showers.  As a result, most protons are only minimum ionizing particles (MIPs) in the LAT, and thus are not useful for the spectrum measurement.  In addition, the event reconstruction is more complicated than for electrons or gamma rays.  The current reconstruction algorithms, optimized for gamma-ray event reconstruction, worked well for the electron analysis.  They do not, however, perform as well in the reconstruction of hadronic events.  For the protons that do produce a shower, the energy resolution is thus much worse than for either gamma rays or electrons, and in fact we have found that the reconstructed energy is approximately a lower limit on the true incoming energy.

In this section, we present our initial findings characterizing the instrument response using a preliminary proton event selection.  We discuss the geometry factor, the background contamination level, and the results from an energy unfolding procedure.  We then conclude with a few notes about future steps in this analysis.

\subsection{Preliminary Geometry Factor}
\begin{figure}[h]
\centering
\includegraphics[width=80mm]{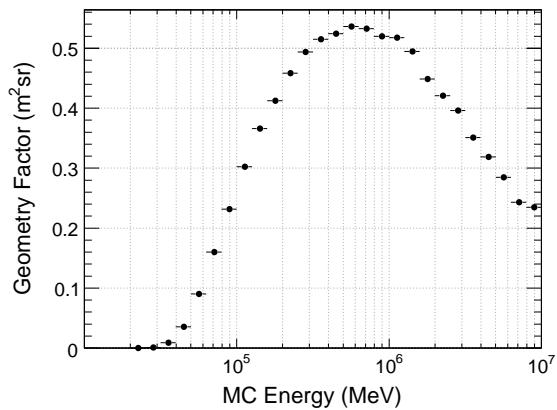}
\caption{The preliminary geometry factor plotted vs. MC energy.} \label{geomfac}
\end{figure}

A preliminary proton event selection has been developed based primarily on an existing classification tree that was designed to distinguish among various particles in the LAT.  For the analysis here, we have also utilized the feature of the onboard filter that allows all events depositing greater than 20 GeV in the calorimeter to be sent to ground without any onboard filtering.  The geometry factor has been calculated from Monte Carlo simulations, with full detector geometry, of protons interacting within the LAT.  A preliminary geometry factor as a function of Monte Carlo (MC) energy, or the true incoming particle energy, is shown in Figure \ref{geomfac}.  It peaks around 0.54 m$^2$sr between 500 and 600 GeV.  The effect of the 20 GeV threshold can be seen as the geometry factor falls to zero near this energy value.  By comparison, the geometry factor for electrons peaks around 2.8 m$^2$sr, which further illustrates the fact that most protons are minimum ionizing in the LAT.  However it is worth noting that this preliminary proton geometry factor is comparable to other cosmic-ray experiments, such as the average value of 0.15 m$^2$sr reported by AMS in 2002 \cite{AMS}.

\subsection{Preliminary Background Contamination}
\begin{figure}[h]
\centering
\includegraphics[width=80mm]{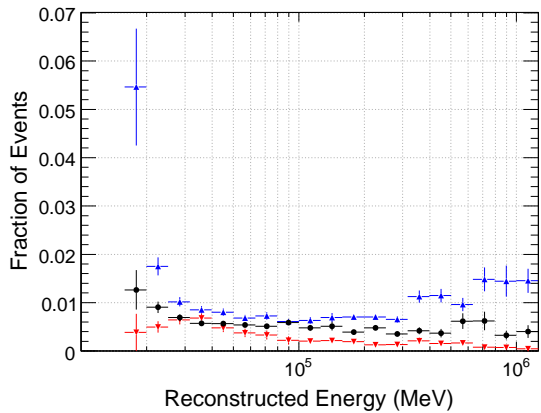}
\caption{The fraction of background contamination for different background components, plotted vs. reconstructed energy, using preliminary proton selections.  Electrons/positrons are plotted in red upside-down triangles, alphas in blue triangles, and particles with Z$>$2 in black circles.  All three components are of order 1\% or less over most of the energy range.} \label{bkgcontam}
\end{figure}

We have estimated the background contamination given these preliminary selections using Fermi Monte Carlo simulations of the cosmic-ray environment the LAT encounters, modelled from previous cosmic-ray measurements, that include the full geometry of the detector.  The dominant background components are found to be electrons/positrons, alphas, and particles with Z$>$2.  For each background component, the fraction of events contained in the final event sample after selections is plotted vs. the reconstructed energy in Figure \ref{bkgcontam}.  Over most of the energy range, each background species contributes only 1\% or less to the selected event sample.

\subsection{Energy Unfolding}
\begin{figure}[h]
\centering
\includegraphics[width=80mm]{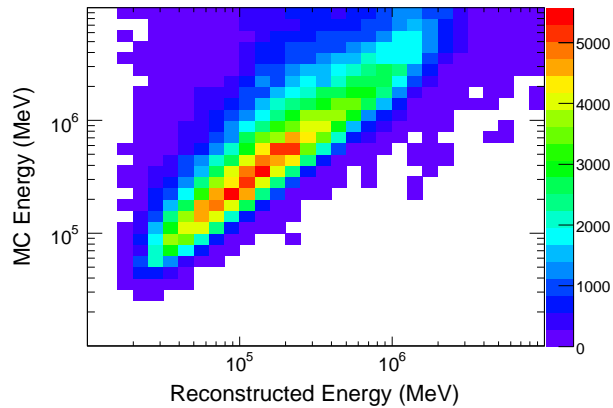}
\caption{The preliminary energy response is plotted as MC energy (y-axis) vs reconstructed energy (x-axis).  For a given bin in MC energy, there is a wide range of reconstructed values obtained.  In addition, the reconstructed value can be seen to be approximately a lower limit on the MC energy.  Both illustrate the challenge in measuring the proton spectrum.} \label{eneresp}
\end{figure}

We have also studied the use of an energy unfolding algorithm on the reconstructed energy distribution, and have studied the instrument's energy response given the preliminary selections.  The process of energy unfolding is to calculate a distribution of the true incoming energies of the events, given a reconstructed energy distribution and a detector response matrix.  It should be noted that this procedure does not attempt to correct energies on an event-by-event basis, but rather its goal is to obtain an estimated distribution for the incoming energies of the selected event population.  The energy unfolding process was also performed for the electron spectrum analysis discussed earlier.  Given that the energy resolution is much worse for protons, the effect of this process on the calculated proton spectrum will be much greater.  The energy response used in the unfolding algorithm is plotted in Figure \ref{eneresp} as MC energy on the y-axis vs. reconstructed energy on the x-axis.  This response has been calculated from Fermi MC proton simulations with a hard spectrum of E$^{-1}$, which allows more events to be generated in the higher energy bins than would the typical cosmic-ray (CR) spectrum, using the preliminary proton selections.  For a given bin in MC energy, there is found to be a wide range of reconstructed energy values, and it can also be seen that the reconstructed energy is approximately a lower limit on the MC energy.

\begin{figure}[h]
\centering
\includegraphics[width=80mm]{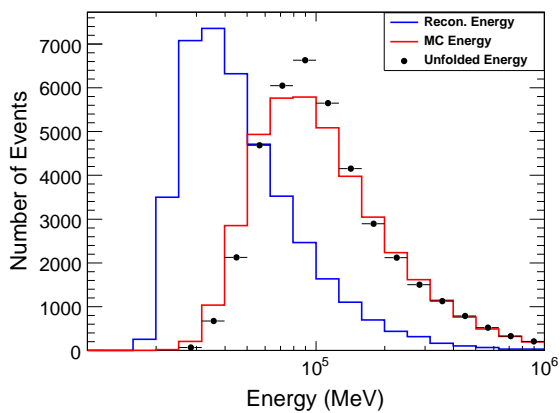}
\caption{Comparison of the unfolded energy distribution (black points) to the MC energy distribution (red line) using preliminary proton selections.  Also shown is the reconstructed energy distribution (blue line) for this selection that was used in the unfolding procedure.} \label{eneunfcomp}
\end{figure}

To perform the unfolding procedure, we have used RooUnfold, a ROOT-based framework for unfolding \cite{RooUnfold}.  The unfolding algorithm takes as input the reconstructed energy distribution and the energy response, and returns the calculated unfolded distribution.  As a test of the unfolding procedure, a comparison can be made between the unfolded energy distribution and the MC energy distribution.  In order make this a fair comparison, the unfolding should be applied to a reconstructed energy distribution taken from a MC sample independent from the one used to create the energy response.  For this test then, we have used the reconstructed energy distribution, after preliminary selections, taken from the Fermi LAT MC cosmic-ray simulations.  The resulting unfolded distribution is shown in Figure \ref{eneunfcomp} along with the MC energy distribution for comparison.  The unfolded distribution obtained is a reasonable reproduction of the MC energy distribution.  However the unfolded distribution underestimates the MC energy distribution at lower energies, while it more closely matches the MC distribution at higher energies.

\subsection{Reconstructed Monte Carlo CR Proton Spectrum}
\begin{figure}[h]
\centering
\includegraphics[width=80mm]{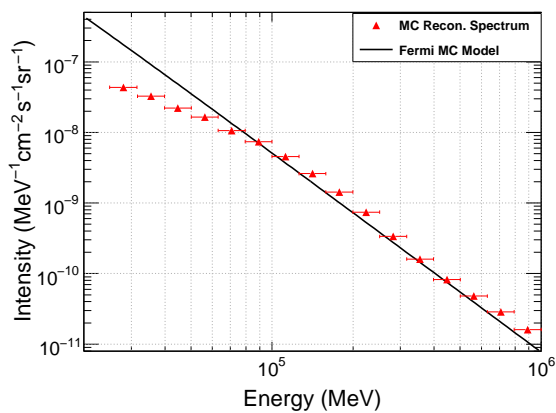}
\caption{The reconstructed MC CR proton spectrum using the unfolded energy distribution (red points).  The spectrum was calculated by dividing the unfolded distribution by the simulated livetime and the geometry factor.  The high-energy behavior of the model used to simulate the MC cosmic-ray proton spectrum is shown for comparison (black line).} \label{reconMCspec}
\end{figure}

The unfolded distribution shown in Figure \ref{eneunfcomp} can be used to calculate the reconstructed MC CR proton spectrum.  The spectrum is calculated by dividing the unfolded distribution by the livetime of the MC simulation used and the geometry factor shown in Figure \ref{geomfac}.  The resulting spectrum is shown in Figure \ref{reconMCspec} with the red triangles.  The solid black line is plotted for comparison and shows the high-energy behavior of the simulated CR proton spectrum, a power law of E$^{-2.83}$.  Similar systematic differences observed between the MC energy distribution and the unfolded energy distribution have been propagated into the spectrum shown here and are evident in the comparison.  In the region below 100 GeV, the reconstructed spectrum turns over and is underestimated.  The spectrum is more closely reconstructed in the region above 100 GeV, however differences are still apparent.

\subsection{Future Steps in Analysis}
Improvements to the preliminary selections used here are being studied and have the goals of spanning a larger energy range and achieving a higher, more constant efficiency and a more constant background contamination level, as a function of energy.  There is also a continuing effort aimed at improving the unfolding procedure to produce a better agreement between the unfolded energy distribution and the MC energy distribution.  With a better understanding of the unfolding procedure and an estimate of the systematic uncertainties, this method can be applied to data collected from the Fermi satellite and the resulting distribution used to reconstruct a proton spectrum.

\bigskip 

\end{document}